\documentclass[amsmath,amssymb,prl,twocolumn,a4paper]{revtex4}
\usepackage{amsfonts} 
\usepackage{amsmath}
\usepackage{graphicx} 
\usepackage{overpic} 
\usepackage{color}

\newcommand{\be}{\begin{eqnarray}} 
\newcommand{\ee}{\end{eqnarray}}
 
\newcommand{\D}{\mathrm{d}}

\begin{document}

\title{Shape universality classes in the random sequential adsorption of non-spherical particles}

\author{Adrian Baule$^1$}

\affiliation{
$^1$School of Mathematical Sciences, Queen Mary University of London, Mile End Road, London E1 4NS, UK
}
\email{a.baule@qmul.ac.uk}

\begin{abstract}

Random sequential adsorption (RSA) of particles of a particular shape is used in a large variety of contexts to model particle aggregation and jamming. A key feature of these models is the observed algebraic time dependence of the asymptotic jamming coverage $\sim t^{-\nu}$ as $t\to\infty$. However, the exact value of the exponent $\nu$ is not known apart from the simplest case of the RSA of monodisperse spheres adsorbed on a line (Renyi's seminal `car parking problem'), where $\nu=1$ can be derived analytically. Empirical simulation studies have conjectured on a case-by-case basis that for general non-spherical particles $\nu=1/(d+\tilde{d})$, where $d$ denotes the dimension of the domain and $\tilde{d}$ the number of orientational degrees of freedom of a particle. Here, we solve this long standing problem analytically for the $d=1$ case --- the `Paris car parking problem'. We prove that the scaling exponent depends on particle shape, contrary to the original conjecture, and, remarkably, falls into two universality classes: (i) $\nu=1/(1+\tilde{d}/2)$ for shapes with a smooth contact distance, e.g., ellipsoids; (ii) $\nu=1/(1+\tilde{d})$ for shapes with a singular contact distance, e.g., spherocylinders and polyhedra. The exact solution explains in particular why many empirically observed scalings fall in between these two limits.

\end{abstract}

\maketitle

The question of how particle shape affects the dynamical and structural properties of particle aggregates is one of the outstanding problems in statistical mechanics with profound technological implications \cite{Onsager:1949aa,Torquato:2002aa,Jaeger:2015aa}. Jammed systems are particularly challenging, since they are dominated by the geometry of the particles and are not described by conventional equilibrium statistical mechanics \cite{Baule:2016aa}. Exploring the effect of shape variation thus relies on extensive computer simulations \cite{Damasceno:2012aa,Miskin:2013aa,Miskin:2014aa,Roth:2016aa} or mean-field theories whose solutions require similar computational efforts \cite{Baule:2013aa,Baule:2014aa}. From a theoretical perspective it is striking that so far there has been hardly any insight from exactly solvable analytical models, even though these are most suitable to identify and classify shapes in the infinite shape space.

In this letter, we consider the probably simplest non-trivial packing model that takes into account excluded volume effects due to shape anisotropies: random sequential adsorption (RSA). Since Renyi's seminal work on the `car parking problem' (the RSA of monodisperse spheres on a line) \cite{Renyi:1958aa,Renyi:1963aa}, RSA models have been widely used to model particle aggregation and jamming in physical, chemical and biological systems \cite{Evans:1993ab,Talbot:2000aa,Cadilhe:2007aa}. Their great appeal is the paradigmatic nature of the adsorption mechanism: the particles' positions and orientations are selected with uniform probability and then placed sequentially into the domain if there is no overlap with any previously placed particles. Particles are not able to move or reorient once being placed. 
 
\begin{figure}
\begin{center}
\includegraphics[width=0.8\columnwidth]{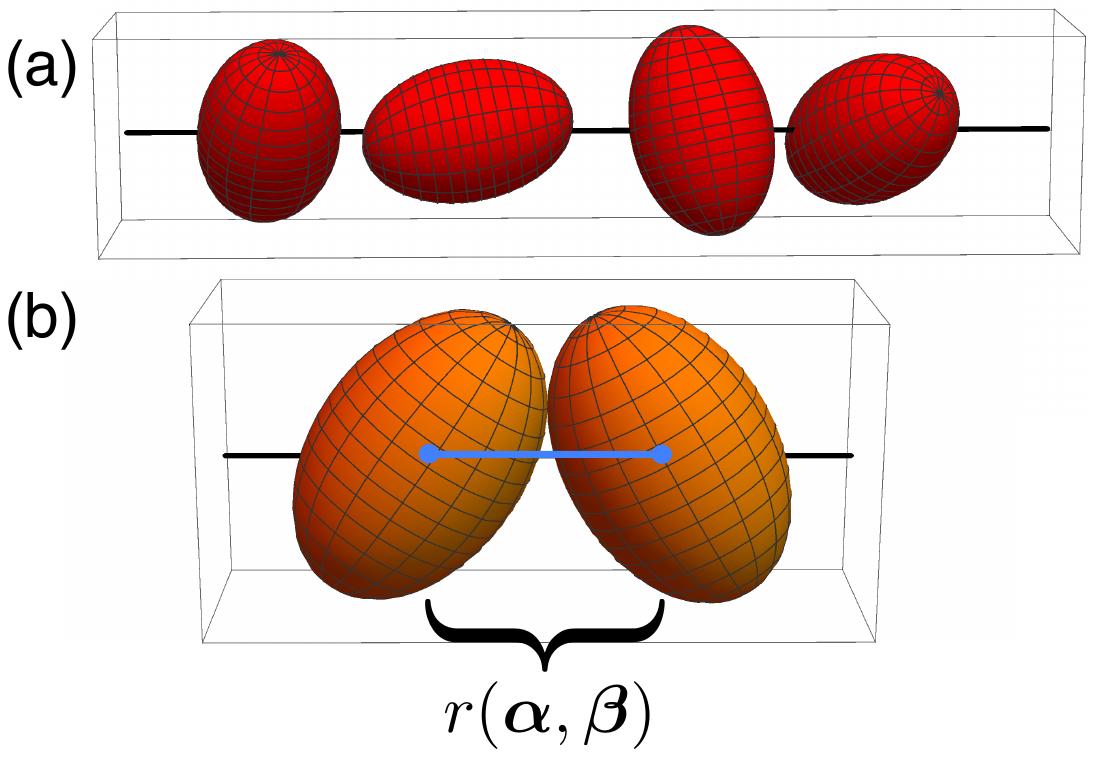}
\caption{\label{figure}(Colors online) (a) Snapshot of a RSA configuration for $d=1$ and $\tilde{d}=2$: a spheroid is selected with a uniform position and orientation. When no overlap with a previously placed particle occurs, it is irreversibly adsorbed on the line. (b) Illustration of the contact distance $r(\boldsymbol\alpha,\boldsymbol\beta)$: the distance of the centres of mass when two particles of orientations $\boldsymbol\alpha,\boldsymbol\beta$ first come into contact. }
\end{center}
\end{figure}
 
Two key features of RSA models are: (i) the existence of a finite jamming density $\phi$ in the infinite time limit $\phi(\infty)=\lim_{t\to\infty}\phi(t)$ and (ii) the algebraic time dependence of the approach to jamming, which has been conjectured as \cite{Viot:1990aa,Evans:1993ab}
\be
\label{conjecture}
\phi(\infty)-\phi(t)\sim t^{-\nu},\qquad \nu=1/(d+\tilde{d})
\ee
for a $d$ dimensional domain and $\tilde{d}$ orientational degrees of freedom of a particle. In the case of spheres ($\tilde{d}=0$), Eq.~\eqref{conjecture} has been initially proposed by Feder \cite{Feder:1980ab} and theoretically supported by Pomeau \cite{Pomeau:1980aa} and Swendsen \cite{Swendsen:1981aa} based on asymptotic estimates. The validity of the conjecture Eq.~\eqref{conjecture} for general non-spherical shapes has been supported from simulation results on a shape-by-shape basis: ellipses \cite{Talbot:1989aa,Sherwood:1990aa,Viot:1992aa}, rectangles \cite{Vigil:1989aa,Vigil:1990aa,Viot:1990aa,Viot:1992aa}, spherocylinders \cite{Viot:1992aa}, and slightly elongated shapes \cite{Ciesla:2014aa}. Approximate theoretical arguments for Eq.~\eqref{conjecture} based on the geometry of target sites in the later stages of the RSA process have been presented in \cite{Talbot:1989aa,Tarjus:1991ac,Viot:1992aa}. Logarithmic corrections have been suggested for cubes \cite{Privman:1991aa}.

Here, we consider the RSA of particles with an arbitrary shape, whose centres of mass fall on a $d=1$ domain (see Fig.~\ref{figure}a). In this case -- referred to as `Paris car parking problem' \cite{Chaikin:2006aa} -- we show below that $\nu$ can be derived in a rigorous way. Remarkably, the exact solution shows that $\nu$ depends not only on $\tilde{d}$ but also on the particle shape manifest in two distinct shape universality classes. Let $p(x,t;\boldsymbol\alpha,\boldsymbol\beta)$ denote the probability to find a segment of length $x$ at time $t$ with a particle of orientation $\boldsymbol\alpha$ at the left boundary of the $x$ interval and of orientation $\boldsymbol\beta$ at the right one. The vector $\boldsymbol\alpha=(\alpha_1,\alpha_2,...,\alpha_{\tilde{d}})$ contains the angles describing the particle's orientation. The master equation for the time evolution of $p$  in dimensionless form is exactly given by
\begin{widetext}
\vspace{-0.5cm}
\be
\label{polyms}
\frac{\partial}{\partial t}p(x,t;\boldsymbol\alpha,\boldsymbol\beta)&=&-\psi(x,\boldsymbol\alpha,\boldsymbol\beta) p(x,t;\boldsymbol\alpha,\boldsymbol\beta)+\left<\int_{x+r(\boldsymbol\beta,\boldsymbol\gamma)}^\infty\D y\,p(y,t;\boldsymbol\alpha,\boldsymbol\gamma)\right>_{\boldsymbol\gamma}+\left<\int_{x+r(\boldsymbol\gamma,\boldsymbol\alpha)}^\infty\D y\,p(y,t;\boldsymbol\gamma,\boldsymbol\beta)\right>_{\boldsymbol\gamma}.
\ee
\end{widetext}

Here, the brackets denote an expected value with respect to the isotropic distribution of the angles: $\left<h(\boldsymbol\gamma)\right>_{\boldsymbol\gamma}=C^{-1}\int\D\boldsymbol\gamma\,h(\boldsymbol\gamma)$, where $C$ is a normalization constant depending on $\tilde{d}$. The function $\psi$ is defined as
\be
\label{psidef}
\psi(x,\boldsymbol\alpha,\boldsymbol\beta)=\left<(x-r(\boldsymbol\alpha,\boldsymbol\gamma)-r(\boldsymbol\gamma,\boldsymbol\beta))^+\right>_{\boldsymbol\gamma}.
\ee
where $(x)^+=x\,\Theta(x)$, i.e., $(x)^+=x$ for $x> 0$ and $(x)^+=0$ for $x\le 0$. The central quantity capturing the effect of anisotropic shapes is $r(\boldsymbol\alpha,\boldsymbol\beta)$ denoting the contact distance of two shapes of orientations $\boldsymbol\alpha$ and $\boldsymbol\beta$ (see Fig.~\ref{figure}b). In Eq.~\eqref{polyms}, the first term on the rhs denotes the probability per unit time that an interval $x,\boldsymbol\alpha,\boldsymbol\beta$ is destroyed by placing a particle inside it ($\psi$ is the probability of insertion). Likewise, the two integrals in Eq.~\eqref{polyms} describe the creation of an interval $x,\boldsymbol\alpha,\boldsymbol\beta$ by placing a particle into a larger interval. 
Eq.~\eqref{polyms} recovers as special cases several models discussed previously in the literature. It trivially recovers the exactly solvable monodisperse sphere case \cite{Renyi:1958aa,Renyi:1963aa,Mackenzie:1962aa,Widom:1966aa}. The next simplest model is the RSA of polydisperse spheres \cite{Ney:1962aa,Mullooly:1968aa,Krapivsky:1992aa,Brilliantov:1996aa,Brilliantov:1997aa,Burridge:2004aa}. Even for this simple extension $\nu$ has not been obtained so far for general size distributions. The special case of bidisperse spheres has been treated in \cite{Hassan:2002aa} showing an algebraic decay with $\nu=1$ due to the small spheres. The $\tilde{d}=1$ version of Eq.~\eqref{polyms} has been studied within an approximate analytical approach in \cite{Tarjus:1991ab} for the case of rectangles in the limit of infinitely long aspect ratios, where $\nu=1/2$ could be confirmed. 

Eq.~\eqref{polyms} separates into three regimes depending on $x$
\be
\label{pcms}
p(x,t;\boldsymbol\alpha,\boldsymbol\beta)=\left\{\begin{matrix}p_1(x,t;\boldsymbol\alpha,\boldsymbol\beta), & x>g_1(\boldsymbol\alpha,\boldsymbol\beta) \\ & \\ p_2(x,t;\boldsymbol\alpha,\boldsymbol\beta), & g_2(\boldsymbol\alpha,\boldsymbol\beta)\le x\le g_1(\boldsymbol\alpha,\boldsymbol\beta) \\ & \\ p_3(x,t;\boldsymbol\alpha,\boldsymbol\beta), & r(\boldsymbol\alpha,\boldsymbol\beta)\le x<g_2(\boldsymbol\alpha,\boldsymbol\beta) \end{matrix}\right.
\ee
In regime 1, $x$ is large enough such that a particle with an arbitrary orientation can be inserted between the two boundary particles. Eq.~\eqref{psidef} then simplifies to
\be
\label{psi1}
\psi(x,\boldsymbol\alpha,\boldsymbol\beta)=x-\left<r(\boldsymbol\alpha,\boldsymbol\gamma)\right>_{\boldsymbol\gamma}-\left<r(\boldsymbol\gamma,\boldsymbol\beta)\right>_{\boldsymbol\gamma}.
\ee
In regime 2, the interval $x$ is not large enough for particles of arbitrary orientations. The constraint on orientations is contained in the full expression Eq.~\eqref{psidef}. In regime 3, $x$ is so small that no particle can be inserted and thus $\psi=0$. The different expressions of $\psi$ are all captured by Eq.~\eqref{psidef}, such that the dynamics in the three regimes is described by Eq.~\eqref{polyms} in a unified way. The three regimes are distinguished by the two functions
\be
g_1(\boldsymbol\alpha,\boldsymbol\beta)&=&\max_{\boldsymbol\gamma}\left[r(\boldsymbol\alpha,\boldsymbol\gamma)+r(\boldsymbol\gamma,\boldsymbol\beta)\right]\\
\label{g1}
g_2(\boldsymbol\alpha,\boldsymbol\beta)&=&\min_{\boldsymbol\gamma}\left[r(\boldsymbol\alpha,\boldsymbol\gamma)+r(\boldsymbol\gamma,\boldsymbol\beta)\right]
\ee
Defining the upper and lower limits of $r$ as $a\le r(\boldsymbol\alpha,\boldsymbol\beta) \le b$, we see that $2a\le g_2(\boldsymbol\alpha,\boldsymbol\beta)\le g_1(\boldsymbol\alpha,\boldsymbol\beta)\le 2b$.

The quantity of main interest in the RSA process is
\be
\label{phit}
\phi(t)=\int\D\boldsymbol\alpha\int\D\boldsymbol\beta\int_{r(\boldsymbol\alpha,\boldsymbol\beta)}^\infty\D x\,p(x,t;\boldsymbol\alpha,\boldsymbol\beta),
\ee
which is the number density of particles, i.e., the 1$d$ equivalent of packing density, which converges to the jamming limit for $t\to\infty$. In order to solve the master equation for $p_1$ we make a similar ansatz as in R\'enyi's car parking problem, which is solved by $p(x,t)=t^2F(t)e^{-xt}$, where $F(t)$ satisfies the ODE $\dot{F}(t)=F(t)\left(a-2\left(1-e^{-at}\right)/t\right)$, assuming spheres of diameter $a$ and the initial condition $F(0)=1$. With Eq.~\eqref{psi1} the ansatz for Eq.~\eqref{polyms} is
\be
\label{p1sol}
p_1(x,t;\boldsymbol\alpha,\boldsymbol\beta)=t^2F(t,\boldsymbol\alpha,\boldsymbol\beta)e^{-xt}
\ee
Substituting into Eq.~(\ref{polyms}) yields 
\be
\label{Falphabeta}
\frac{\partial}{\partial t}F(t,\boldsymbol\alpha,\boldsymbol\beta)&=&(\left<r(\boldsymbol\alpha,\boldsymbol\gamma)\right>_{\boldsymbol\gamma}+\left<r(\boldsymbol\gamma,\boldsymbol\beta)\right>_{\boldsymbol\gamma})F(t,\boldsymbol\alpha,\boldsymbol\beta)\nonumber\\
&&-\frac{2F(t,\boldsymbol\alpha,\boldsymbol\beta)}{t}+\frac{1}{t}\left<F(t,\boldsymbol\alpha,\boldsymbol\gamma)e^{-r(\boldsymbol\beta,\boldsymbol\gamma)t}\right>_{\boldsymbol\gamma}\nonumber\\
&&+\frac{1}{t}\left<F(t,\boldsymbol\gamma,\boldsymbol\beta)e^{-r(\boldsymbol\gamma,\boldsymbol\alpha)t}\right>_{\boldsymbol\gamma}.
\ee
The key observation is that also the master equation for $p_2$ can be solved analytically for moderate aspect ratios of the particles. We define the length scale $\tilde{g}(\boldsymbol\alpha,\boldsymbol\beta)=\min_{\gamma}\left[r(\boldsymbol\alpha,\boldsymbol\gamma)+g_2(\boldsymbol\gamma,\boldsymbol\beta)\right]=\min_{\gamma}\left[g_2(\boldsymbol\alpha,\boldsymbol\gamma)+r(\boldsymbol\gamma,\boldsymbol\beta)\right]$. The length $\tilde{g}$ is interpreted as follows: for $g_2\le x\le \tilde{g}$, $x$ is so small that maximally one particle can be placed inside it. This means that if $\tilde{g}\ge g_1$ for all $\boldsymbol\alpha,\boldsymbol\beta$ the dynamics in regime 2 simplifies: the integral terms in Eq.~(\ref{polyms}) only integrate over $p_1$. As a result $p_2$ satisfies a simple first-order ODE with an inhomogeneity. The solution is
\be
\label{p2sol}
p_2(x,t;\boldsymbol\alpha,\boldsymbol\beta)&=&p_1(x,t;\boldsymbol\alpha,\boldsymbol\beta)+\left[x-\left<r(\boldsymbol\alpha,\boldsymbol\gamma)\right>_{\boldsymbol\gamma}\right.\nonumber\\
&&-\left.\left<r(\boldsymbol\gamma,\boldsymbol\beta)\right>_{\boldsymbol\gamma}-\psi(x,\boldsymbol\alpha,\boldsymbol\beta)\right]\\
&&\times\int_0^t\D se^{-\psi(x,\boldsymbol\alpha,\boldsymbol\beta)(t-s)}p_1(x,s;\boldsymbol\alpha,\boldsymbol\beta).\nonumber
\ee
Crucially, the condition $\tilde{g}\ge g_1$ is satisfied for $3a\ge 2b$, since then $\tilde{g}\ge 3a\ge 2b \ge g_1$ for all $\boldsymbol\alpha,\boldsymbol\beta$. For regular convex particles, $a,b$ can be identified with the width and length of the particles, respectively, so $p_2$ is given by Eq.~\eqref{p2sol} for aspect ratios $\le 3/2$. Since we can express $p_3$ analytically with the Eqs.~(\ref{p1sol},\ref{p2sol}) (the integral terms in the master equation for $p_3$ only contain $p_{1,2}$), the only remaining unknown is the function $F$ of Eq.~\eqref{Falphabeta} in $p_1$.

The contact distance is in general a highly complicated function, which already in the case of ellipsoids can not be expressed in closed form \cite{Zheng:2009aa}. Solving Eq.~\eqref{Falphabeta} analytically is thus not feasible in general. However, the exponent $\nu$ can still be determined. We need to calculate
\be
\label{asympstep}
&&\phi(\infty)-\phi(t)\nonumber\\
&&=\int\D\boldsymbol\alpha\int\D\boldsymbol\beta\int_{r(\boldsymbol\alpha,\boldsymbol\beta)}^\infty\D x\int_t^\infty\D s\frac{\partial p}{\partial s}(x,s;\boldsymbol\alpha,\boldsymbol\beta).
\ee
Substituting the master equation for the time derivative in each of the $x$ regimes, we see that we need to evaluate time integrals of $p_{1,2}$. The key to express these analytically is that $F$ scales for large $t$ as (from Eq.~\eqref{Falphabeta})
\be
\label{Fscaling}
F(t,\boldsymbol\alpha,\boldsymbol\beta)&\approx& e^{\left(\left<r(\boldsymbol\alpha,\boldsymbol\gamma)\right>_{\boldsymbol\gamma}+\left<r(\boldsymbol\gamma,\boldsymbol\beta)\right>_{\boldsymbol\gamma}\right)(t-t_{\rm c})-2\int_{t_{\rm c}}^t\D s\,\frac{1}{s}}\nonumber\\
&\sim& t^{-2}e^{\left(\left<r(\boldsymbol\alpha,\boldsymbol\gamma)\right>_{\boldsymbol\gamma}+\left<r(\boldsymbol\gamma,\boldsymbol\beta)\right>_{\boldsymbol\gamma}\right)t},
\ee
where $t_c$ is a lower cutoff of order one that does not contribute to the asymptotic scaling. With Eq.~\eqref{Fscaling} the asymptotics of the integrals over $p_{1,2}$ can also be determined. From Eqs.~(\ref{asympstep},\ref{Fscaling}) we obtain with some manipulations \cite{SM}
\be
\label{phit}
&&\phi(\infty)-\phi(t)\sim\int\D\boldsymbol\alpha\int\D\boldsymbol\beta\int_{g_2(\boldsymbol\alpha,\boldsymbol\beta)}^{g_1(\boldsymbol\alpha,\boldsymbol\beta)}\D x\,e^{-\psi(x,\boldsymbol\alpha,\boldsymbol\beta)t}\nonumber\\
&&+\int\D\boldsymbol\alpha\int\D\boldsymbol\beta\int_{g_1(\boldsymbol\alpha,\boldsymbol\beta)}^\infty\D x\, e^{-(x-\left<r(\boldsymbol\alpha,\boldsymbol\gamma)\right>_{\boldsymbol\gamma}-\left<r(\boldsymbol\gamma,\boldsymbol\beta)\right>_{\boldsymbol\gamma})t}
\ee
Since $\left<r(\boldsymbol\alpha,\boldsymbol\gamma)\right>_{\boldsymbol\gamma}+\left<r(\boldsymbol\gamma,\boldsymbol\beta)\right>_{\boldsymbol\gamma}>g_1(\boldsymbol\alpha,\boldsymbol\beta)$, the second term decays exponentially for $t\to\infty$. The asymptotic scaling is thus determined by evaluating the asymptotics of the first Laplace-type integral. To this end we need to investigate the stationary points of $\psi$. The definitions of $\psi$ and $g_2$ imply that $\psi(g_2(\boldsymbol\alpha,\boldsymbol\beta),\boldsymbol\alpha,\boldsymbol\beta)=0$. Calculating the gradient of $\psi$, we also obtain $\nabla\psi(g_2(\boldsymbol\alpha,\boldsymbol\beta),\boldsymbol\alpha,\boldsymbol\beta)=0$, so the stationary points lie on the surface $x=g_2(\boldsymbol\alpha,\boldsymbol\beta)$ on the boundary of the integration region and correspond to minima since $\psi\ge 0$. The asymptotics of such a high-dimensional Laplace integral with degenerate stationary points is typically highly challenging. The analysis in the present case is possible since the behaviour of $\psi$ for $x$ close to the minima can be determined analytically. Using Eq.~\eqref{psidef} we can write
\be
\label{psi2}
\psi(x,\boldsymbol\alpha,\boldsymbol\beta)=\frac{1}{C}\sum_{i=1}^n\int_{\Omega_i}\D\boldsymbol\gamma(x-r(\boldsymbol\alpha,\boldsymbol\gamma)-r(\boldsymbol\gamma,\boldsymbol\beta)),
\ee
where it is assumed that there are $n$ $\tilde{d}$-dimensional domains $\Omega_i(x,\boldsymbol\alpha,\boldsymbol\beta)$ where $\psi\ge 0$, i.e., $\Omega_i$ is bounded by hypersurfaces satisfying
\be
\label{xgammai}
x= r(\boldsymbol\alpha,\boldsymbol\gamma)+r(\boldsymbol\gamma,\boldsymbol\beta).
\ee
We first assume a unique global minimum $\boldsymbol\gamma^*(\boldsymbol\alpha,\boldsymbol\beta)$ for all configurations $\boldsymbol\alpha,\boldsymbol\beta$ corresponding to the $\boldsymbol\gamma$ value defining $g_2(\boldsymbol\alpha,\boldsymbol\beta)$ in Eq.~\eqref{g1}. As $x\to g_2(\boldsymbol\alpha,\boldsymbol\beta)$ only the interval $i^*$ containing $\boldsymbol\gamma^*$ remains in the sum in Eq.~\eqref{psi2}. Expanding around $\boldsymbol\gamma^*$ thus yields to leading order
\be
\label{psi3}
\psi(x,\boldsymbol\alpha,\boldsymbol\beta)\approx\frac{1}{C}(x-g_2(\boldsymbol\alpha,\boldsymbol\beta))\Omega_{i^*}(x,\boldsymbol\alpha,\boldsymbol\beta).
\ee
The volume $\Omega_{i^*}$ is centred at $\boldsymbol\gamma^*$ and constrained to become smaller and smaller for $x\to g_2(\boldsymbol\alpha,\boldsymbol\beta)$. If we introduce the vector $\boldsymbol\epsilon=\boldsymbol\gamma-\boldsymbol\gamma^*$ and switch to spherical coordinates $\boldsymbol\epsilon=z(\boldsymbol\theta)\mathbf{\hat{u}}({\boldsymbol\theta})$, where $\boldsymbol\theta$ parametrizes the solid angle in $\tilde{d}$ dimensions and $\mathbf{\hat{u}}$ is a unit vector, we can calculate $\Omega_{i^*}$ as $\Omega_{i^*}=\oint\D\boldsymbol\theta\int_0^{z(\boldsymbol\theta)}\D z\,z^{\tilde{d}-1}=\frac{1}{\tilde{d}}\oint\D\boldsymbol\theta \,z(\boldsymbol\theta)^{\tilde{d}}$, where $z(\boldsymbol\theta)$ denotes the boundary of the volume $\Omega_{i^*}$ in the direction of a given solid angle $\boldsymbol\theta$ and $\D\boldsymbol\theta$ includes the surface element in $\tilde{d}$ dimensions. This means that $z(\boldsymbol\theta)=z(\boldsymbol\theta;x,\boldsymbol\alpha,\boldsymbol\beta)$ and is determined by the condition Eq.~\eqref{xgammai}. In order to determine $z$, we develop Eq.~\eqref{xgammai} around $\boldsymbol\gamma^*$. This yields up to quadratic orders $x\approx g_2(\boldsymbol\alpha,\boldsymbol\beta)+\boldsymbol\epsilon^{\rm T}{\rm M}\boldsymbol\epsilon$, where ${\rm M}(\boldsymbol\alpha,\boldsymbol\beta)=\nabla_{\boldsymbol\gamma}\nabla_{\boldsymbol\gamma} r(\boldsymbol\alpha,\boldsymbol\gamma^*)+\nabla_{\boldsymbol\gamma}\nabla_{\boldsymbol\gamma} r(\boldsymbol\gamma^*,\boldsymbol\beta)$. As a consequence $z(\boldsymbol\theta;x,\boldsymbol\alpha,\boldsymbol\beta)$ is given by $z=\sqrt{x-g_2(\boldsymbol\alpha,\boldsymbol\beta)}/(\mathbf{\hat{u}}({\boldsymbol\theta})^{\rm T}{\rm M}(\boldsymbol\alpha,\boldsymbol\beta)\mathbf{\hat{u}}({\boldsymbol\theta}))$ and the leading order of $\psi$ is with Eqs.~(\ref{psi3})
\be
\psi(x,\boldsymbol\alpha,\boldsymbol\beta)\approx\oint\D\boldsymbol\theta\frac{(x-g_2(\boldsymbol\alpha,\boldsymbol\beta))^{1+\tilde{d}/2}}{C\left(\mathbf{\hat{u}}({\boldsymbol\theta})^{\rm T}{\rm M}(\boldsymbol\alpha,\boldsymbol\beta)\mathbf{\hat{u}}({\boldsymbol\theta})\right)^{\tilde{d}}}.
\ee
For large $t$, Eq.~\eqref{phit} yields after a variable transformation
\be
\label{scaling1}
\phi(\infty)-\phi(t)\sim\int_0^{1}\D x\,e^{-x^{1+\tilde{d}/2}t}\sim t^{-1/(1+\tilde{d}/2)}
\ee
where the upper limit of the $x$ integration is irrelevant since both $g_1(\boldsymbol\alpha,\boldsymbol\beta)-g_2(\boldsymbol\alpha,\boldsymbol\beta)$ and $\oint\D\boldsymbol\theta \left(\mathbf{\hat{u}}({\boldsymbol\theta})^{\rm T}{\rm M}(\boldsymbol\alpha,\boldsymbol\beta)\mathbf{\hat{u}}({\boldsymbol\theta})\right)^{-\tilde{d}/(1+\tilde{d}/2)}$ are finite of order one. Note, however, that the minimum $\boldsymbol\gamma^*$ can be continuously degenerate for a range of configurations $\boldsymbol\alpha,\boldsymbol\beta$ depending on the shape. For spheroids and spherocylinders, e.g., when both $\boldsymbol\alpha,\boldsymbol\beta$ are perpendicular to the axis, one can rotate a particle placed in between at contact without changing the contact distances. For each degeneracy with respect to a finite rotation, the effective dimensionality is reduced by one, since the volume $\Omega_{i^*}$ does not shrink in one of the directions as $x\to g_2(\boldsymbol\alpha,\boldsymbol\beta)$. These contributions to the asymptotic scaling are subdominant since they decay as $t^{-1/(1+(\tilde{d}-l)/2)}$ for $l$ such degeneracies and thus Eq.~\eqref{scaling1} prevails as $t\to\infty$.

\begin{figure}
\begin{center}
\includegraphics[width=7.5cm]{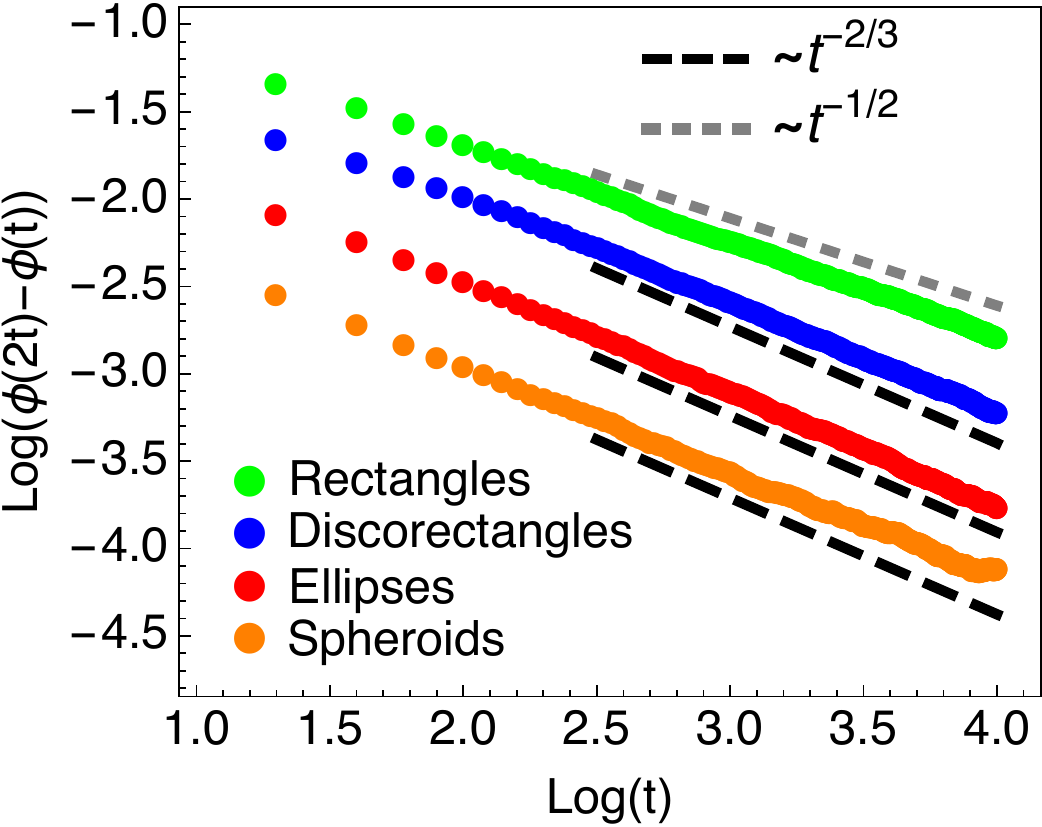}
\caption{\label{Fig_scaling}(Colors online) Plot of simulation results for the asymptotic scaling for a set of shapes with aspect ratio 1.5 \cite{SM}. Shown is the function $\log(\phi(2t)-\phi(t))$, which exhibits the same scaling as $\log(\phi(\infty)-\phi(t))$ when plotted against $\log(t)$ \cite{Vigil:1990aa}. For rectangles, discorectangles, and spheroids the empirical exponent falls in the range $1/2\le\nu\le2/3$ indicating an intermediate time regime as explained by the theory. Data for $\tilde{d}=1$ ($\tilde{d}=2$) shapes are averaged over 500 (200) samples.}
\end{center}
\end{figure}

The scaling Eq.~\eqref{scaling1} holds when $r$ is smooth in every direction around the minimum $\boldsymbol\gamma^*$. This is true for any smooth convex shape with non-zero curvature. On the other hand, if the shape has sections of flat sides the expansion up to quadratic order breaks down, since the minimum $\boldsymbol\gamma^*$ can be singular. In order to elucidate the situation, we consider first the $\tilde{d}=1$ case, where $r$ can be approximated in closed analytical form for small angles $\alpha,\beta$ \cite{Lebowitz:1987aa,Kantor:2009aa}
\be
\label{rapprox}
r(\alpha,\beta)\approx 2a+a_1(\alpha^2+\beta^2)+a_2|\alpha-\beta|^\mu,
\ee 
Here, $a_{1,2}$ and $\mu$ are shape dependent parameters. For generic shapes, $\mu$ is given by either $1$ or $2$ depending on whether the contact point is away from the axis or close to it, respectively. For ellipses, $\mu=2$ and $r$ is smooth throughout. For rectangles, $\mu=1$ and $r$ is singular when the minimum is at $\gamma^*=\alpha$ or $\gamma^*=\beta$. In arbitrary dimensions, we can infer from Eq.~\eqref{rapprox} that the singular behaviour around $\boldsymbol\gamma^*$ is likewise governed by an absolute value in one or multiple directions for shapes with flat sides \cite{Lebowitz:1987aa}. The integration region $\Omega_{i^*}$ can then be separated in a piecewise way and close to $\boldsymbol\gamma^*$ be expanded up to linear order in each of the regions: $x\approx g_2(\boldsymbol\alpha,\boldsymbol\beta)+\mathbf{h}^{(j)}\,\boldsymbol\epsilon$ with $\mathbf{h}(\boldsymbol\alpha,\boldsymbol\beta)={\nabla_{\boldsymbol\gamma} r}(\boldsymbol\alpha,\boldsymbol\gamma^*)+{\nabla r_{\boldsymbol\gamma}}(\boldsymbol\gamma^*,\boldsymbol\beta)$. Since the first order term $\mathbf{h}^{(j)}$ of the $j$th region does not vanish, we have $z^{(j)}=(x-g_2(\boldsymbol\alpha,\boldsymbol\beta))/(\mathbf{h}^{(j)}(\boldsymbol\alpha,\boldsymbol\beta)\mathbf{\hat{u}}(\boldsymbol\theta))$. The leading term of $\psi$ in this case is
\be
\psi(x,\boldsymbol\alpha,\boldsymbol\beta)\approx\sum_{j=1}^m\int_j\D\boldsymbol\theta\frac{(x-g_2(\boldsymbol\alpha,\boldsymbol\beta))^{1+\tilde{d}}}{C\left(\mathbf{h}^{(j)}(\boldsymbol\alpha,\boldsymbol\beta)\mathbf{\hat{u}}(\boldsymbol\theta)\right)^{\tilde{d}}},
\ee
assuming $m$ piecewise regions of the integration domain covering different solid angles. The asymptotic scaling is then for arbitrary dimensions
\be
\label{scaling2}
\phi(\infty)-\phi(t)\sim\int_0^{1}\D x\,e^{-x^{1+\tilde{d}}t}\sim t^{-1/(1+\tilde{d})}.
\ee
Importantly, the singular nature of $\boldsymbol\gamma^*$ varies depending on $\boldsymbol\alpha,\boldsymbol\beta$. For rectangles and discorectangles under the approximation Eq.~\eqref{rapprox}, there are many configurations where $\gamma^*\neq \alpha,\beta$ and the minimum is smooth. In general, the corresponding regions in $\boldsymbol\alpha,\boldsymbol\beta$ need to be separated in the integral Eq.~\eqref{phit}. The overall scaling is then given as a superposition of terms proportional to $t^{-1/(1+\tilde{d}/2)}$ and $t^{-1/(1+\tilde{d})}$. As $t\to\infty$ the $t^{-1/(1+\tilde{d})}$ scaling always dominates, but this might be visible only on very long time scales.

Comparing the theoretical predictions Eqs.~(\ref{scaling1},\ref{scaling2}) with simulation data, we see that the scaling $t^{-2/3}$ for ellipses ($\tilde{d}=1$) is clearly observed (see Fig.~\ref{Fig_scaling}). The data for rectangles and discorectangles lies in between the predicted $t^{-2/3}$ and $t^{-1/2}$ scalings indicating an intermediate time regime, since the minimum can be both singular and smooth depending on $\alpha,\beta$. In $\tilde{d}=2$, the solution predicts the scaling $\nu= 1/2$ for spheroids. This scaling is not observed on the time scales accessible in the simulations of Fig.~\ref{Fig_scaling}. The reason is that apart from specific configurations leading to degenerate minima, there exist also quasi-degeneracies for almost all $\boldsymbol\alpha,\boldsymbol\beta$ reducing the effective dimensionality by one before the $t^{-1/2}$ scaling is attained for very long times \cite{SM}. In Fig.~\ref{Fig_scaling} the spheroid data indeed shows an intermediate $t^{-2/3}$ scaling over a considerable range. The quasi degeneracies are due to the short aspect ratio regime and reduced for larger aspect ratios, where small angular differences can induce more pronounced variations in the contact distance \cite{SM}.

The results Eqs.~(\ref{scaling1},\ref{scaling2}) are rigorous for particles with aspect ratio $\le 3/2$. However, the same results are expected to hold for arbitrary aspect ratios, since the asymptotic scaling in the RSA process will be dominated by the filling of the smallest $x$ intervals, in which particles can still be placed. These are intervals $g_2\le x\le \tilde{g}$, such that the corresponding $p$ will always decay as the solution of Eq.~\eqref{p2sol} for large $t$ and the same analysis holds.

In summary, the analytical solution of the `Paris car parking problem' solves a long standing problem in our understanding of RSA processes highlighting the breakdown of the conjecture Eq.~\eqref{conjecture} and connecting the scaling exponent directly with shape features. The analysis of the function $r(\boldsymbol\alpha,\boldsymbol\gamma)+r(\boldsymbol\gamma,\boldsymbol\beta)$ shows the existence of two shape universality classes depending on the presence of singularities at the minimum $\gamma^*$. The exact geometry of target sites thus intimately affects the kinetics, which should also be true in higher dimensions. It would be very interesting to find out if the same or similar universality classes govern also other jamming properties for non-spherical shapes, e.g., the observed peak in the packing density at specific aspect ratios of elongated shapes (see, e.g., \cite{Williams:2003aa,Donev:2004aa,Baule:2013aa,Ciesla:2016aa}), which allows the identification of optimally dense granular packings that are highly relevant for developing new functional granular materials \cite{Jaeger:2015aa}. Since the model Eq.~\eqref{polyms} captures the exact hard core excluded volume of shapes and exhibits a density peak as shown in simulations of ellipses \cite{Chaikin:2006aa}, an analytical analysis of the peak in this model would be feasible if the jamming density $\phi(\infty)$ could be calculated. In turn, this requires the explicit solution of Eq.~\eqref{Falphabeta}, which will be investigated in the future.

The results highlight the importance of a precise modelling of the particle shape, since even small shape differences can lead to rather distinct kinetics for large times. Such an insight is important, e.g., to improve the modelling of nucleosome adsorption on DNA, which is described by variants of RSA processes on a 1$d$ line \cite{Ranjith:2007aa,Padinhateeri:2011aa,Osberg:2014aa}. Nucleosomes indeed have non-spherical shapes and can adsorb in variable orientations \cite{Fritzsche:1996aa,Funke:2016aa}. Models that incorporate these degrees of freedom could thus provide valuable new insight.

\begin{acknowledgments}

AB gratefully acknowledges funding under EPSRC grant EP/L020955/1 and thanks O.~Bandtlow, A.~Gnedin, and W.~Just for helpful discussions.

\end{acknowledgments}


\begin{thebibliography}{99}
\expandafter\ifx\csname natexlab\endcsname\relax\def\natexlab#1{#1}\fi
\expandafter\ifx\csname bibnamefont\endcsname\relax
  \def\bibnamefont#1{#1}\fi
\expandafter\ifx\csname bibfnamefont\endcsname\relax
  \def\bibfnamefont#1{#1}\fi
\expandafter\ifx\csname citenamefont\endcsname\relax
  \def\citenamefont#1{#1}\fi
\expandafter\ifx\csname url\endcsname\relax
  \def\url#1{\texttt{#1}}\fi
\expandafter\ifx\csname urlprefix\endcsname\relax\def\urlprefix{URL }\fi
\providecommand{\bibinfo}[2]{#2}
\providecommand{\eprint}[2][]{\url{#2}}

\bibitem[{\citenamefont{Onsager}(1949)}]{Onsager:1949aa}
\bibinfo{author}{\bibfnamefont{L.}~\bibnamefont{Onsager}},
  \bibinfo{journal}{Ann. N. Y. Acad. Sci.} \textbf{\bibinfo{volume}{51}},
  \bibinfo{pages}{627} (\bibinfo{year}{1949}).

\bibitem[{\citenamefont{Torquato}(2002)}]{Torquato:2002aa}
\bibinfo{author}{\bibfnamefont{S.}~\bibnamefont{Torquato}},
  \emph{\bibinfo{title}{Random heterogeneous materials: microstructure and
  macroscopic properties}} (\bibinfo{publisher}{Springer},
  \bibinfo{year}{2002}).

\bibitem[{\citenamefont{Jaeger}(2015)}]{Jaeger:2015aa}
\bibinfo{author}{\bibfnamefont{H.~M.} \bibnamefont{Jaeger}},
  \bibinfo{journal}{Soft Matter} \textbf{\bibinfo{volume}{11}},
  \bibinfo{pages}{12} (\bibinfo{year}{2015}).

\bibitem[{\citenamefont{{Baule} et~al.}(2016)\citenamefont{{Baule}, {Morone},
  {Herrmann}, and {Makse}}}]{Baule:2016aa}
\bibinfo{author}{\bibfnamefont{A.}~\bibnamefont{{Baule}}},
  \bibinfo{author}{\bibfnamefont{F.}~\bibnamefont{{Morone}}},
  \bibinfo{author}{\bibfnamefont{H.~J.} \bibnamefont{{Herrmann}}},
  \bibnamefont{and} \bibinfo{author}{\bibfnamefont{H.~A.}
  \bibnamefont{{Makse}}}, \bibinfo{journal}{ArXiv e-prints}
  (\bibinfo{year}{2016}), \eprint{1602.04369}.

\bibitem[{\citenamefont{Damasceno et~al.}(2012)\citenamefont{Damasceno, Engel,
  and Glotzer}}]{Damasceno:2012aa}
\bibinfo{author}{\bibfnamefont{P.~F.} \bibnamefont{Damasceno}},
  \bibinfo{author}{\bibfnamefont{M.}~\bibnamefont{Engel}}, \bibnamefont{and}
  \bibinfo{author}{\bibfnamefont{S.~C.} \bibnamefont{Glotzer}},
  \bibinfo{journal}{Science} \textbf{\bibinfo{volume}{337}},
  \bibinfo{pages}{453} (\bibinfo{year}{2012}).

\bibitem[{\citenamefont{Miskin and Jaeger}(2013)}]{Miskin:2013aa}
\bibinfo{author}{\bibfnamefont{M.~Z.} \bibnamefont{Miskin}} \bibnamefont{and}
  \bibinfo{author}{\bibfnamefont{H.~M.} \bibnamefont{Jaeger}},
  \bibinfo{journal}{Nature Mater.} \textbf{\bibinfo{volume}{12}},
  \bibinfo{pages}{326} (\bibinfo{year}{2013}).

\bibitem[{\citenamefont{Miskin and Jaeger}(2014)}]{Miskin:2014aa}
\bibinfo{author}{\bibfnamefont{M.~Z.} \bibnamefont{Miskin}} \bibnamefont{and}
  \bibinfo{author}{\bibfnamefont{H.~M.} \bibnamefont{Jaeger}},
  \bibinfo{journal}{Soft Matter} \textbf{\bibinfo{volume}{10}},
  \bibinfo{pages}{3708} (\bibinfo{year}{2014}).

\bibitem[{\citenamefont{Roth and Jaeger}(2016)}]{Roth:2016aa}
\bibinfo{author}{\bibfnamefont{L.~K.} \bibnamefont{Roth}} \bibnamefont{and}
  \bibinfo{author}{\bibfnamefont{H.~M.} \bibnamefont{Jaeger}},
  \bibinfo{journal}{Soft Matter} \textbf{\bibinfo{volume}{12}},
  \bibinfo{pages}{1107} (\bibinfo{year}{2016}).

\bibitem[{\citenamefont{Baule et~al.}(2013)\citenamefont{Baule, Mari, Bo,
  Portal, and Makse}}]{Baule:2013aa}
\bibinfo{author}{\bibfnamefont{A.}~\bibnamefont{Baule}},
  \bibinfo{author}{\bibfnamefont{R.}~\bibnamefont{Mari}},
  \bibinfo{author}{\bibfnamefont{L.}~\bibnamefont{Bo}},
  \bibinfo{author}{\bibfnamefont{L.}~\bibnamefont{Portal}}, \bibnamefont{and}
  \bibinfo{author}{\bibfnamefont{H.~A.} \bibnamefont{Makse}},
  \bibinfo{journal}{Nature Commun.} \textbf{\bibinfo{volume}{4}},
  \bibinfo{pages}{2194} (\bibinfo{year}{2013}).

\bibitem[{\citenamefont{Baule and Makse}(2014)}]{Baule:2014aa}
\bibinfo{author}{\bibfnamefont{A.}~\bibnamefont{Baule}} \bibnamefont{and}
  \bibinfo{author}{\bibfnamefont{H.~A.} \bibnamefont{Makse}},
  \bibinfo{journal}{Soft Matter} \textbf{\bibinfo{volume}{10}},
  \bibinfo{pages}{4423} (\bibinfo{year}{2014}).

\bibitem[{\citenamefont{R\'enyi}(1958)}]{Renyi:1958aa}
\bibinfo{author}{\bibfnamefont{A.}~\bibnamefont{R\'enyi}},
  \bibinfo{journal}{Publ. Math. Res. Inst. Hung. Acad. Sci.}
  \textbf{\bibinfo{volume}{3}}, \bibinfo{pages}{109} (\bibinfo{year}{1958}).

\bibitem[{\citenamefont{Renyi}(1963)}]{Renyi:1963aa}
\bibinfo{author}{\bibfnamefont{A.}~\bibnamefont{R\'enyi}}, \bibinfo{journal}{Sel.
  Trans. Math. Stat. Prob.} \textbf{\bibinfo{volume}{4}}, \bibinfo{pages}{205}
  (\bibinfo{year}{1963}).

\bibitem[{\citenamefont{Evans}(1993)}]{Evans:1993ab}
\bibinfo{author}{\bibfnamefont{J.~W.} \bibnamefont{Evans}},
  \bibinfo{journal}{Rev. Mod. Phys.} \textbf{\bibinfo{volume}{65}},
  \bibinfo{pages}{1281} (\bibinfo{year}{1993}).

\bibitem[{\citenamefont{Talbot et~al.}(2000)\citenamefont{Talbot, Tarjus,
  Tassel, and Viot}}]{Talbot:2000aa}
\bibinfo{author}{\bibfnamefont{J.}~\bibnamefont{Talbot}},
  \bibinfo{author}{\bibfnamefont{G.}~\bibnamefont{Tarjus}},
  \bibinfo{author}{\bibfnamefont{P.~V.} \bibnamefont{Tassel}},
  \bibnamefont{and} \bibinfo{author}{\bibfnamefont{P.}~\bibnamefont{Viot}},
  \bibinfo{journal}{Colloids Surf., A} \textbf{\bibinfo{volume}{165}}, \bibinfo{pages}{287 }
  (\bibinfo{year}{2000}).

\bibitem[{\citenamefont{Cadilhe et~al.}(2007)\citenamefont{Cadilhe, Ara{\'u}jo,
  and Privman}}]{Cadilhe:2007aa}
\bibinfo{author}{\bibfnamefont{A.}~\bibnamefont{Cadilhe}},
  \bibinfo{author}{\bibfnamefont{N.~A.~M.} \bibnamefont{Ara{\'u}jo}},
  \bibnamefont{and} \bibinfo{author}{\bibfnamefont{V.}~\bibnamefont{Privman}},
  \bibinfo{journal}{J. Phys. Condens. Matter}
  \textbf{\bibinfo{volume}{19}}, \bibinfo{pages}{065124}
  (\bibinfo{year}{2007}).

\bibitem[{\citenamefont{Viot and Tarjus}(1990)}]{Viot:1990aa}
\bibinfo{author}{\bibfnamefont{P.}~\bibnamefont{Viot}} \bibnamefont{and}
  \bibinfo{author}{\bibfnamefont{G.}~\bibnamefont{Tarjus}},
  \bibinfo{journal}{Europhys. Lett.} \textbf{\bibinfo{volume}{13}},
  \bibinfo{pages}{295} (\bibinfo{year}{1990}).

\bibitem[{\citenamefont{Feder}(1980)}]{Feder:1980ab}
\bibinfo{author}{\bibfnamefont{J.}~\bibnamefont{Feder}},
  \bibinfo{journal}{J. Theor. Biol.}
  \textbf{\bibinfo{volume}{87}}, \bibinfo{pages}{237 } (\bibinfo{year}{1980}),
  ISSN \bibinfo{issn}{0022-5193}.

\bibitem[{\citenamefont{Pomeau}(1980)}]{Pomeau:1980aa}
\bibinfo{author}{\bibfnamefont{Y.}~\bibnamefont{Pomeau}},
  \bibinfo{journal}{J. Phys. A}
  \textbf{\bibinfo{volume}{13}}, \bibinfo{pages}{L193} (\bibinfo{year}{1980}).

\bibitem[{\citenamefont{Swendsen}(1981)}]{Swendsen:1981aa}
\bibinfo{author}{\bibfnamefont{R.~H.} \bibnamefont{Swendsen}},
  \bibinfo{journal}{Phys. Rev. A} \textbf{\bibinfo{volume}{24}},
  \bibinfo{pages}{504} (\bibinfo{year}{1981}).

\bibitem[{\citenamefont{Talbot et~al.}(1989)\citenamefont{Talbot, Tarjus, and
  Schaaf}}]{Talbot:1989aa}
\bibinfo{author}{\bibfnamefont{J.}~\bibnamefont{Talbot}},
  \bibinfo{author}{\bibfnamefont{G.}~\bibnamefont{Tarjus}}, \bibnamefont{and}
  \bibinfo{author}{\bibfnamefont{P.}~\bibnamefont{Schaaf}},
  \bibinfo{journal}{Phys. Rev. A} \textbf{\bibinfo{volume}{40}},
  \bibinfo{pages}{4808} (\bibinfo{year}{1989}).

\bibitem[{\citenamefont{Sherwood}(1990)}]{Sherwood:1990aa}
\bibinfo{author}{\bibfnamefont{J.~D.} \bibnamefont{Sherwood}},
  \bibinfo{journal}{J. Phys. A}
  \textbf{\bibinfo{volume}{23}}, \bibinfo{pages}{2827} (\bibinfo{year}{1990}).

\bibitem[{\citenamefont{Viot et~al.}(1992)\citenamefont{Viot, Tarjus, Ricci,
  and Talbot}}]{Viot:1992aa}
\bibinfo{author}{\bibfnamefont{P.}~\bibnamefont{Viot}},
  \bibinfo{author}{\bibfnamefont{G.}~\bibnamefont{Tarjus}},
  \bibinfo{author}{\bibfnamefont{S.~M.} \bibnamefont{Ricci}}, \bibnamefont{and}
  \bibinfo{author}{\bibfnamefont{J.}~\bibnamefont{Talbot}},
  \bibinfo{journal}{J. Chem. Phys.} \textbf{\bibinfo{volume}{97}},
  \bibinfo{pages}{5212} (\bibinfo{year}{1992}).

\bibitem[{\citenamefont{Vigil and Ziff}(1989)}]{Vigil:1989aa}
\bibinfo{author}{\bibfnamefont{R.~D.} \bibnamefont{Vigil}} \bibnamefont{and}
  \bibinfo{author}{\bibfnamefont{R.~M.} \bibnamefont{Ziff}},
  \bibinfo{journal}{J. Chem. Phys.} \textbf{\bibinfo{volume}{91}},
  \bibinfo{pages}{2599} (\bibinfo{year}{1989}).

\bibitem[{\citenamefont{Vigil and Ziff}(1990)}]{Vigil:1990aa}
\bibinfo{author}{\bibfnamefont{R.~D.} \bibnamefont{Vigil}} \bibnamefont{and}
  \bibinfo{author}{\bibfnamefont{R.~M.} \bibnamefont{Ziff}},
  \bibinfo{journal}{J. Chem. Phys.} \textbf{\bibinfo{volume}{93}},
  \bibinfo{pages}{8270} (\bibinfo{year}{1990}).

\bibitem[{\citenamefont{Ciesla and Barbasz}(2014)}]{Ciesla:2014aa}
\bibinfo{author}{\bibfnamefont{M.}~\bibnamefont{Ciesla}} \bibnamefont{and}
  \bibinfo{author}{\bibfnamefont{J.}~\bibnamefont{Barbasz}},
  \bibinfo{journal}{Phys. Rev. E} \textbf{\bibinfo{volume}{89}},
  \bibinfo{pages}{022401} (\bibinfo{year}{2014}).

\bibitem[{\citenamefont{Tarjus and Talbot}(1991)}]{Tarjus:1991ac}
\bibinfo{author}{\bibfnamefont{G.}~\bibnamefont{Tarjus}} \bibnamefont{and}
  \bibinfo{author}{\bibfnamefont{J.}~\bibnamefont{Talbot}},
  \bibinfo{journal}{J. Phys. A}
  \textbf{\bibinfo{volume}{24}}, \bibinfo{pages}{L913} (\bibinfo{year}{1991}).

\bibitem[{\citenamefont{Privman et~al.}(1991)\citenamefont{Privman, Wang, and
  Nielaba}}]{Privman:1991aa}
\bibinfo{author}{\bibfnamefont{V.}~\bibnamefont{Privman}},
  \bibinfo{author}{\bibfnamefont{J.-S.} \bibnamefont{Wang}}, \bibnamefont{and}
  \bibinfo{author}{\bibfnamefont{P.}~\bibnamefont{Nielaba}},
  \bibinfo{journal}{Phys. Rev. B} \textbf{\bibinfo{volume}{43}},
  \bibinfo{pages}{3366} (\bibinfo{year}{1991}).

\bibitem[{\citenamefont{Chaikin et~al.}(2006)\citenamefont{Chaikin, Donev, Man,
  Stillinger, and Torquato}}]{Chaikin:2006aa}
\bibinfo{author}{\bibfnamefont{P.~M.} \bibnamefont{Chaikin}},
  \bibinfo{author}{\bibfnamefont{A.}~\bibnamefont{Donev}},
  \bibinfo{author}{\bibfnamefont{W.}~\bibnamefont{Man}},
  \bibinfo{author}{\bibfnamefont{F.~H.} \bibnamefont{Stillinger}},
  \bibnamefont{and} \bibinfo{author}{\bibfnamefont{S.}~\bibnamefont{Torquato}},
  \bibinfo{journal}{Ind. Eng. Chem. Res.} \textbf{\bibinfo{volume}{45}},
  \bibinfo{pages}{6960} (\bibinfo{year}{2006}).

\bibitem[{\citenamefont{Mackenzie}(1962)}]{Mackenzie:1962aa}
\bibinfo{author}{\bibfnamefont{J.~K.} \bibnamefont{Mackenzie}},
  \bibinfo{journal}{J. Chem. Phys.}
  \textbf{\bibinfo{volume}{37}}, \bibinfo{pages}{723} (\bibinfo{year}{1962}).

\bibitem[{\citenamefont{Widom}(1966)}]{Widom:1966aa}
\bibinfo{author}{\bibfnamefont{B.}~\bibnamefont{Widom}}, \bibinfo{journal}{J.
  Chem. Phys.} \textbf{\bibinfo{volume}{44}}, \bibinfo{pages}{3888}
  (\bibinfo{year}{1966}).

\bibitem[{\citenamefont{Ney}(1962)}]{Ney:1962aa}
\bibinfo{author}{\bibfnamefont{P.~E.} \bibnamefont{Ney}},
  \bibinfo{journal}{Ann. Math. Stat.} \textbf{\bibinfo{volume}{33}},
  \bibinfo{pages}{702} (\bibinfo{year}{1962}).

\bibitem[{\citenamefont{Mullooly}(1968)}]{Mullooly:1968aa}
\bibinfo{author}{\bibfnamefont{J.~P.} \bibnamefont{Mullooly}},
  \bibinfo{journal}{J. Appl. Probab.}
  \textbf{\bibinfo{volume}{5}}, \bibinfo{pages}{427} (\bibinfo{year}{1968}).

\bibitem[{\citenamefont{Krapivsky}(1992)}]{Krapivsky:1992aa}
\bibinfo{author}{\bibfnamefont{P.~L.} \bibnamefont{Krapivsky}},
  \bibinfo{journal}{J. Stat. Phys.}
  \textbf{\bibinfo{volume}{69}}, \bibinfo{pages}{135} (\bibinfo{year}{1992}).

\bibitem[{\citenamefont{Brilliantov et~al.}(1996)\citenamefont{Brilliantov,
  Andrienko, Krapivsky, and Kurths}}]{Brilliantov:1996aa}
\bibinfo{author}{\bibfnamefont{N.~V.} \bibnamefont{Brilliantov}},
  \bibinfo{author}{\bibfnamefont{Y.~A.} \bibnamefont{Andrienko}},
  \bibinfo{author}{\bibfnamefont{P.~L.} \bibnamefont{Krapivsky}},
  \bibnamefont{and} \bibinfo{author}{\bibfnamefont{J.}~\bibnamefont{Kurths}},
  \bibinfo{journal}{Phys. Rev. Lett.} \textbf{\bibinfo{volume}{76}},
  \bibinfo{pages}{4058} (\bibinfo{year}{1996}).

\bibitem[{\citenamefont{Brilliantov et~al.}(1997)\citenamefont{Brilliantov,
  Andrienko, and Krapivsky}}]{Brilliantov:1997aa}
\bibinfo{author}{\bibfnamefont{N.}~\bibnamefont{Brilliantov}},
  \bibinfo{author}{\bibfnamefont{Y.}~\bibnamefont{Andrienko}},
  \bibnamefont{and}
  \bibinfo{author}{\bibfnamefont{P.}~\bibnamefont{Krapivsky}},
  \bibinfo{journal}{Physica A}
  \textbf{\bibinfo{volume}{239}}, \bibinfo{pages}{267 } (\bibinfo{year}{1997}).

\bibitem[{\citenamefont{Burridge and Mao}(2004)}]{Burridge:2004aa}
\bibinfo{author}{\bibfnamefont{D.~J.}~\bibnamefont{Burridge}} \bibnamefont{and}
  \bibinfo{author}{\bibfnamefont{Y.}~\bibnamefont{Mao}},
  \bibinfo{journal}{Phys. Rev. E} \textbf{\bibinfo{volume}{69}},
  \bibinfo{pages}{037102} (\bibinfo{year}{2004}).

\bibitem[{\citenamefont{Hassan et~al.}(2002)\citenamefont{Hassan, Schmidt,
  Blasius, and Kurths}}]{Hassan:2002aa}
\bibinfo{author}{\bibfnamefont{M.~K.} \bibnamefont{Hassan}},
  \bibinfo{author}{\bibfnamefont{J.}~\bibnamefont{Schmidt}},
  \bibinfo{author}{\bibfnamefont{B.}~\bibnamefont{Blasius}}, \bibnamefont{and}
  \bibinfo{author}{\bibfnamefont{J.}~\bibnamefont{Kurths}},
  \bibinfo{journal}{Phys. Rev. E} \textbf{\bibinfo{volume}{65}},
  \bibinfo{pages}{045103} (\bibinfo{year}{2002}).

\bibitem[{\citenamefont{Tarjus and Viot}(1991)}]{Tarjus:1991ab}
\bibinfo{author}{\bibfnamefont{G.}~\bibnamefont{Tarjus}} \bibnamefont{and}
  \bibinfo{author}{\bibfnamefont{P.}~\bibnamefont{Viot}},
  \bibinfo{journal}{Phys. Rev. Lett.} \textbf{\bibinfo{volume}{67}},
  \bibinfo{pages}{1875} (\bibinfo{year}{1991}).

\bibitem[{\citenamefont{Zheng et~al.}(2009)\citenamefont{Zheng, Iglesias, and
  Palffy-Muhoray}}]{Zheng:2009aa}
\bibinfo{author}{\bibfnamefont{X.}~\bibnamefont{Zheng}},
  \bibinfo{author}{\bibfnamefont{W.}~\bibnamefont{Iglesias}}, \bibnamefont{and}
  \bibinfo{author}{\bibfnamefont{P.}~\bibnamefont{Palffy-Muhoray}},
  \bibinfo{journal}{Phys. Rev. E} \textbf{\bibinfo{volume}{79}},
  \bibinfo{pages}{057702} (\bibinfo{year}{2009}).

\bibitem[{\citenamefont{Lebowitz et~al.}(1987)\citenamefont{Lebowitz, Percus,
  and Talbot}}]{Lebowitz:1987aa}
\bibinfo{author}{\bibfnamefont{J.~L.} \bibnamefont{Lebowitz}},
  \bibinfo{author}{\bibfnamefont{J.~K.} \bibnamefont{Percus}},
  \bibnamefont{and} \bibinfo{author}{\bibfnamefont{J.}~\bibnamefont{Talbot}},
  \bibinfo{journal}{J. Stat. Phys.}
  \textbf{\bibinfo{volume}{49}}, \bibinfo{pages}{1221} (\bibinfo{year}{1987}),
  ISSN \bibinfo{issn}{1572-9613}.

\bibitem[{\citenamefont{Kantor and Kardar}(2009)}]{Kantor:2009aa}
\bibinfo{author}{\bibfnamefont{Y.}~\bibnamefont{Kantor}} \bibnamefont{and}
  \bibinfo{author}{\bibfnamefont{M.}~\bibnamefont{Kardar}},
  \bibinfo{journal}{Europhys. Lett.} \textbf{\bibinfo{volume}{87}},
  \bibinfo{pages}{60002} (\bibinfo{year}{2009}).

\bibitem[{\citenamefont{Williams and Philipse}(2003)}]{Williams:2003aa}
\bibinfo{author}{\bibfnamefont{S.~R.} \bibnamefont{Williams}} \bibnamefont{and}
  \bibinfo{author}{\bibfnamefont{A.~P.} \bibnamefont{Philipse}},
  \bibinfo{journal}{Phys. Rev. E} \textbf{\bibinfo{volume}{67}},
  \bibinfo{pages}{051301} (\bibinfo{year}{2003}).

\bibitem[{\citenamefont{Donev et~al.}(2004)\citenamefont{Donev, Cisse, Sachs,
  Variano, Stillinger, Connelly, Torquato, and Chaikin}}]{Donev:2004aa}
\bibinfo{author}{\bibfnamefont{A.}~\bibnamefont{Donev}},
  \bibinfo{author}{\bibfnamefont{I.}~\bibnamefont{Cisse}},
  \bibinfo{author}{\bibfnamefont{D.}~\bibnamefont{Sachs}},
  \bibinfo{author}{\bibfnamefont{E.}~\bibnamefont{Variano}},
  \bibinfo{author}{\bibfnamefont{F.}~\bibnamefont{Stillinger}},
  \bibinfo{author}{\bibfnamefont{R.}~\bibnamefont{Connelly}},
  \bibinfo{author}{\bibfnamefont{S.}~\bibnamefont{Torquato}}, \bibnamefont{and}
  \bibinfo{author}{\bibfnamefont{P.}~\bibnamefont{Chaikin}},
  \bibinfo{journal}{Science} \textbf{\bibinfo{volume}{303}},
  \bibinfo{pages}{990} (\bibinfo{year}{2004}).

\bibitem[{\citenamefont{Cie{\'s}la et~al.}(2016)\citenamefont{Cie{\'s}la,
  Pajak, and Ziff}}]{Ciesla:2016aa}
\bibinfo{author}{\bibfnamefont{M.}~\bibnamefont{Cie{\'s}la}},
  \bibinfo{author}{\bibfnamefont{G.}~\bibnamefont{Pajak}}, \bibnamefont{and}
  \bibinfo{author}{\bibfnamefont{R.~M.} \bibnamefont{Ziff}},
  \bibinfo{journal}{J. Chem. Phys.}
  \textbf{\bibinfo{volume}{145}}, \bibinfo{pages}{044708}
  (\bibinfo{year}{2016}).

\bibitem[{\citenamefont{Ranjith et~al.}(2007)\citenamefont{Ranjith, Yan, and
  Marko}}]{Ranjith:2007aa}
\bibinfo{author}{\bibfnamefont{P.}~\bibnamefont{Ranjith}},
  \bibinfo{author}{\bibfnamefont{J.}~\bibnamefont{Yan}}, \bibnamefont{and}
  \bibinfo{author}{\bibfnamefont{J.~F.} \bibnamefont{Marko}},
  \bibinfo{journal}{Proc. Natl. Acad. Sci. U.S.A.}
  \textbf{\bibinfo{volume}{104}}, \bibinfo{pages}{13649}
  (\bibinfo{year}{2007}).

\bibitem[{\citenamefont{Padinhateeri and Marko}(2011)}]{Padinhateeri:2011aa}
\bibinfo{author}{\bibfnamefont{R.}~\bibnamefont{Padinhateeri}}
  \bibnamefont{and} \bibinfo{author}{\bibfnamefont{J.~F.} \bibnamefont{Marko}},
  \bibinfo{journal}{Proc. Natl. Acad. Sci. U.S.A.}
  \textbf{\bibinfo{volume}{108}}, \bibinfo{pages}{7799} (\bibinfo{year}{2011}).

\bibitem[{\citenamefont{Osberg et~al.}(2014)\citenamefont{Osberg, Nuebler,
  Korber, and Gerland}}]{Osberg:2014aa}
\bibinfo{author}{\bibfnamefont{B.}~\bibnamefont{Osberg}},
  \bibinfo{author}{\bibfnamefont{J.}~\bibnamefont{Nuebler}},
  \bibinfo{author}{\bibfnamefont{P.}~\bibnamefont{Korber}}, \bibnamefont{and}
  \bibinfo{author}{\bibfnamefont{U.}~\bibnamefont{Gerland}},
  \bibinfo{journal}{Nucleic Acids Res.} \textbf{\bibinfo{volume}{42}},
  \bibinfo{pages}{13633} (\bibinfo{year}{2014}).

\bibitem[{\citenamefont{Fritzsche and Henderson}(1996)}]{Fritzsche:1996aa}
\bibinfo{author}{\bibfnamefont{W.}~\bibnamefont{Fritzsche}} \bibnamefont{and}
  \bibinfo{author}{\bibfnamefont{E.}~\bibnamefont{Henderson}},
  \bibinfo{journal}{Biophys. J.} \textbf{\bibinfo{volume}{71}},
  \bibinfo{pages}{2222 } (\bibinfo{year}{1996}).

\bibitem[{\citenamefont{Funke et~al.}(2016)\citenamefont{Funke, Ketterer,
  Lieleg, Schunter, Korber, and Dietz}}]{Funke:2016aa}
\bibinfo{author}{\bibfnamefont{J.~J.} \bibnamefont{Funke}},
  \bibinfo{author}{\bibfnamefont{P.}~\bibnamefont{Ketterer}},
  \bibinfo{author}{\bibfnamefont{C.}~\bibnamefont{Lieleg}},
  \bibinfo{author}{\bibfnamefont{S.}~\bibnamefont{Schunter}},
  \bibinfo{author}{\bibfnamefont{P.}~\bibnamefont{Korber}}, \bibnamefont{and}
  \bibinfo{author}{\bibfnamefont{H.}~\bibnamefont{Dietz}},
  \bibinfo{journal}{Sci. Adv.} \textbf{\bibinfo{volume}{2}}
  (\bibinfo{year}{2016}).


\bibitem{SM}See Supplemental Material [url] for more details, which includes Ref.~\cite{Abreu:2003aa}.


\bibitem{Abreu:2003aa}
\bibinfo{author}{\bibfnamefont{C.}~\bibnamefont{Abreu}},
  \bibinfo{author}{\bibfnamefont{F.}~\bibnamefont{Tavares}}, \bibnamefont{and}
  \bibinfo{author}{\bibfnamefont{M.}~\bibnamefont{Castier}},
  \bibinfo{journal}{Powder Technol.} \textbf{\bibinfo{volume}{134}},
  \bibinfo{pages}{167} (\bibinfo{year}{2003}).


\end{thebibliography}
\end{document}